# Understanding spectral artefacts in SKA-Low 21-cm cosmology experiments: the impact of cable reflections

Oscar S. D. O'Hara[1,2]★ Fred Dulwich,[1,2] Eloy de Lera Acedo,[1,2] Jiten Dhandha[2,3], Thomas Gessey-Jones[1,2], Dominic Anstey[1,2] and Anastasia Fialkov[2,3]

[1]*Astrophysics Group, Cavendish Astrophysics, J.J. Thomson Avenue, Cambridge, CB3 0HE, UK*
[2]*Kavli Institute for Cosmology in Cambridge, Madingley Road, Cambridge, CB3 0HA, UK*
[3]*Institute of Astronomy, University of Cambridge, Madingley Road, Cambridge, CB3 0HA, UK*

## ABSTRACT

The Cosmic Dawn was marked by the formation of the first stars, and preceded the Epoch of Reionization (EoR), when the Universe underwent a fundamental transformation caused by the radiation from these first stars and galaxies. Interferometric 21-cm experiments aim to probe redshifted neutral hydrogen signals from these periods, constraining the conditions of the early Universe. The SKA-Low instrument of the Square Kilometre Array (SKA) is envisaged to be the largest and most sensitive radio telescope at metre and centimetre wavelengths. The latest Aperture Array Verification Systems feature 7-m coaxial transmission lines connecting the low noise amplifiers to optical transmitters at the front of the analogue-receiving chain. An impedance mismatch between these components results in a partially reflected electromagnetic signal, introducing chromatic aberrations into the instrument bandpass. This causes power from the foreground signals to appear at higher delays, potentially contaminating the 'EoR window', a region in which the 21-cm signal should be detectable. We present an end-to-end simulation pipeline for SKA-Low using a composite sky model combining radio foregrounds from the Galactic and Extragalactic All-Sky MWA (GLEAM) Survey, Haslam 408 MHz, and a 1.5-cGpc 21-cm brightness temperature cube generated with the 21cmSPACE simulator. We derive a model for the scattering parameters of a coaxial transmission line in terms of its specifications and bulk material properties. Assuming identical cables of length $\leq 15.0$ m with impedance mismatch $\leq 10\,\Omega$, the reflection is confined below the EoR window. However, we demonstrate that even a 0.1 per cent length tolerance introduces contamination with an absolute fractional difference of ∼10 per cent across all accessible k-modes.

## 1 INTRODUCTION

The measurement of spatial fluctuations of the 21-cm cosmological signal is one of the primary scientific goals of the Square Kilometre Array (SKA) radio telescope, and is crucial to understanding the development of the Universe from the Dark Ages to the Epoch of Reionization (EoR; Furlanetto, Oh & Briggs 2006). In the early Universe, statistical inhomogeneities in the distribution of matter and the local astrophysical properties of galaxies, combined with the cosmological light-cone effect, result in a spectrally varying 21-cm power spectrum (Matsubara, Suto & Szapudi 1997). Theoretical models predict that the sensitivity of the SKA-Low will be sufficient to detect the 21-cm power spectrum at $k = [0.1, 0.33]$ Mpc$^{-1}$ (Cohen, Fialkov & Barkana 2018; Reis, Fialkov & Barkana 2020; Barkana et al. 2023).

However, the primary obstacle preventing such a detection is the isolation of the 21-cm signal from that of the radio frequency (RF) foreground emission, which is predominantly composed of spectrally smooth Galactic synchrotron emission and free–free emission (Shaver et al. 1999) whose components extend up to five orders of magnitude greater than the 21-cm signal (Furlanetto 2016). These components can be separated based on their spectral composition in delay space, defined as the Fourier-transformed frequency axis of the visibility spectrum (the 'natural' measurement of an interferometer).

The EoR window refers to a delay space region relatively free of smooth spectral contamination such as foregrounds (Vedantham, Shankar & Subrahmanyan 2012; Thyagarajan et al. 2013, 2015, 2016; Liu, Parsons & Trott 2014a, b; Trott & Wayth 2016). The spectral composition of a propagating signal is distorted by chromaticity associated with instrumental systematics. In the case of 21-cm cosmology, smooth radio foregrounds are pushed beyond the band-limited foreground wedge, contaminating the EoR window as shown in Fig. 1.

To date, there exist two types of 21-cm cosmological experiments, differentiated by instrument design. The first type is the global 21-cm experiments, which aim to detect the spatially averaged 21-cm signal. There are several examples of such experiments, such as the Experiment to Detect the Global EoR Signature (EDGES; Bowman, Rogers & Hewitt 2008), the Shaped Antenna measurement of the background RAdio Spectrum (SARAS; Patra et al. 2013), the Mapper of the IGM Spin Temperature (MIST; Monsalve et al. 2024), the Large-aperture

★ E-mail: osdo2@cam.ac.uk

Experiment to detect the Dark Age (LEDA; Price et al. 2018) and the Radio Experiment for the Analysis of Cosmic Hydrogen (REACH; de Lera Acedo et al. 2022). The second type of experiment, and the focal point of this paper, is the use of interferometric instruments to detect the full spatially varying power spectrum of the 21-cm signal. Some examples of this include the LOw-Frequency ARray (LOFAR; van Haarlem et al. 2013), the Hydrogen Epoch of Reionization Array (HERA; DeBoer et al. 2017), the Precision Array for Probing the Epoch of Re-ionization (PAPER; Parsons et al. 2010), the Murchison Widefield Array (MWA; Tingay et al. 2013), the New Extension in Nançay Upgrading LOFAR (NenUFAR; Zarka et al. 2015), the Long Wavelength Array at the Owens Valley Radio Observatory (OVRO-LWA; Hallinan 2015) and what will be the Square Kilometre Array (SKA; Dewdney et al. 2009).

The second Aperture Array Verification System (AAVS2; Van Es et al. 2020; Wayth et al. 2022) is an engineering platform array composed of 256 Square Kilometre Array Log-periodic Antenna (SKALA-4.1) elements (Bolli et al. 2020). The SKALA-4.1 implemented within AAVS2 is the most recent development within the SKALA family (de Lera Acedo 2012; de Lera Acedo et al. 2015a, b), designed to maximize sensitivity across a $\pm 45°$ field of view, while maintaining high polarization purity and a flat impedance across the SKA-Low band. AAVS2 features a 7-m coaxial transmission line between each antenna element and one of the 16 Small Modular Aggregation and RFoF Trunk (SMART) boxes, where the RF signal is converted to optical wavelengths for its long-range transmission (Perini et al. 2022).

Transmission line theory predicts that a propagating electromagnetic (EM) signal may experience a partial reflection (reflection coefficient $\Gamma$) at a boundary as a result of an impedance mismatch between the input (component characteristic impedance $Z_0$) and output (load impedance $Z_L$) domains (Ellingson, Walz & Browder 2018; Ellingson 2020),

$$\Gamma = \frac{Z_L - Z_0}{Z_L + Z_0}. \tag{1}$$

Regardless of the quality of construction and component design, an impedance mismatch will exist between the output of the antenna low-noise amplifier (LNA), the connecting coaxial transmission line, and the input of the SMART box. These cable reflections introduce additional chromatic structure in the instrument passband in the form of a decaying sinusoid. The time-domain reflection structure can be observed by taking the Fourier transform, revealing the contaminating power and the associated delays.

This chromatic structure is further complicated by transmission line manufacturing tolerances, on-site environmental conditions, and the potential damage the cable may sustain during construction (Kruger et al. 2009). The implementation of either foreground avoidance (HERA) or a foreground removal (MWA, LOFAR, and NenUFAR) paradigm requires a fundamental understanding of foreground structure; however, these reflections are known to propagate the foreground power to higher delays (Kern et al. 2020), deviating from existing calibration models, and potentially contaminating the EoR window. In this work, we recognize the importance of having general calibration measures and quality assurance metrics to effectively remove contaminating systematics and capture the 21-cm cosmological signal. However, implementing these would necessitate a separate pipeline, so we have deferred this aspect for the future.

In this paper, we present an end-to-end simulation pipeline for the quantification of SKA-Low spectral artefacts. Section 2 outlines the various pipeline components required to simulate and analyse the interferometric visibilities in the presence of a systematic effect. In Section 3, the EoR window is divided into two distinct k-mode regions: those containing suppressed foreground emission and those dominated by supra-horizon emission. Behaviour-dependent figures of merit provide a robust and confident quantitative assessment of foreground contamination and are applied across the transmission line parameter space in Section 4. Finally, in Section 5, we summarize our main conclusions.

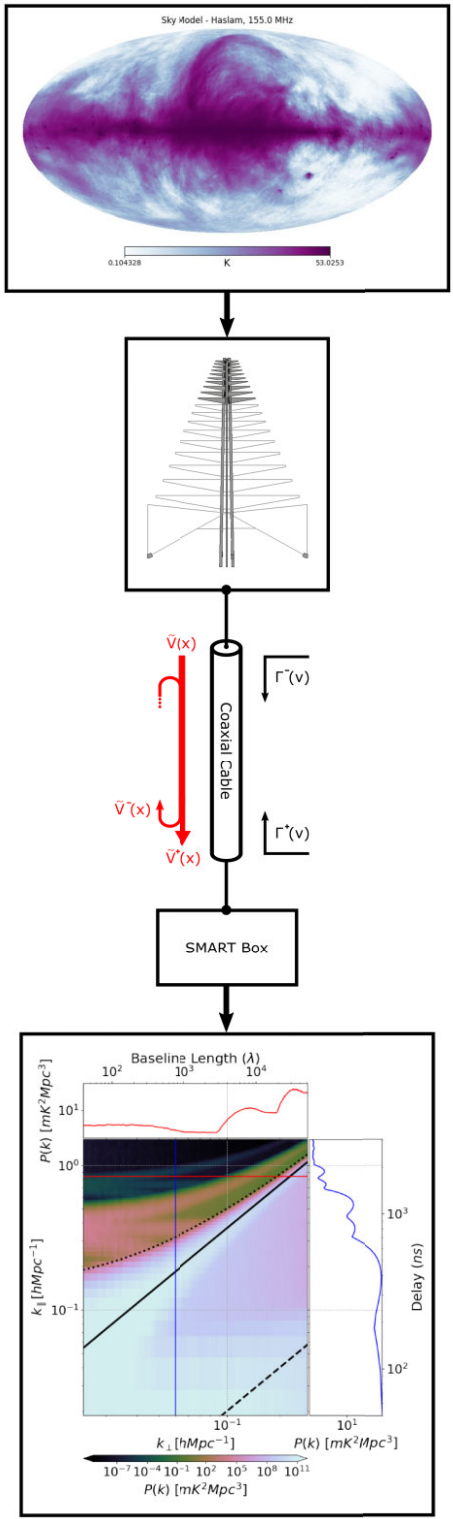

**Figure 1.** For SKA-Low, the interferometric response of each SKALA element will propagate from the LNA as an electrical signal along a coaxial transmission line towards the SMART boxes for long-range transmission to the beamformer. If an impedance mismatch exists between the front-end components and the transmission line, the signal will experience a partial reflection. As a result of these reflections, smooth radio foregrounds typically confined below the horizon limit are pushed to higher delays, contaminating the EoR window. The delay power spectrum displayed corresponds to the foreground contamination effect observed for a 100-m RG58u coaxial cable.

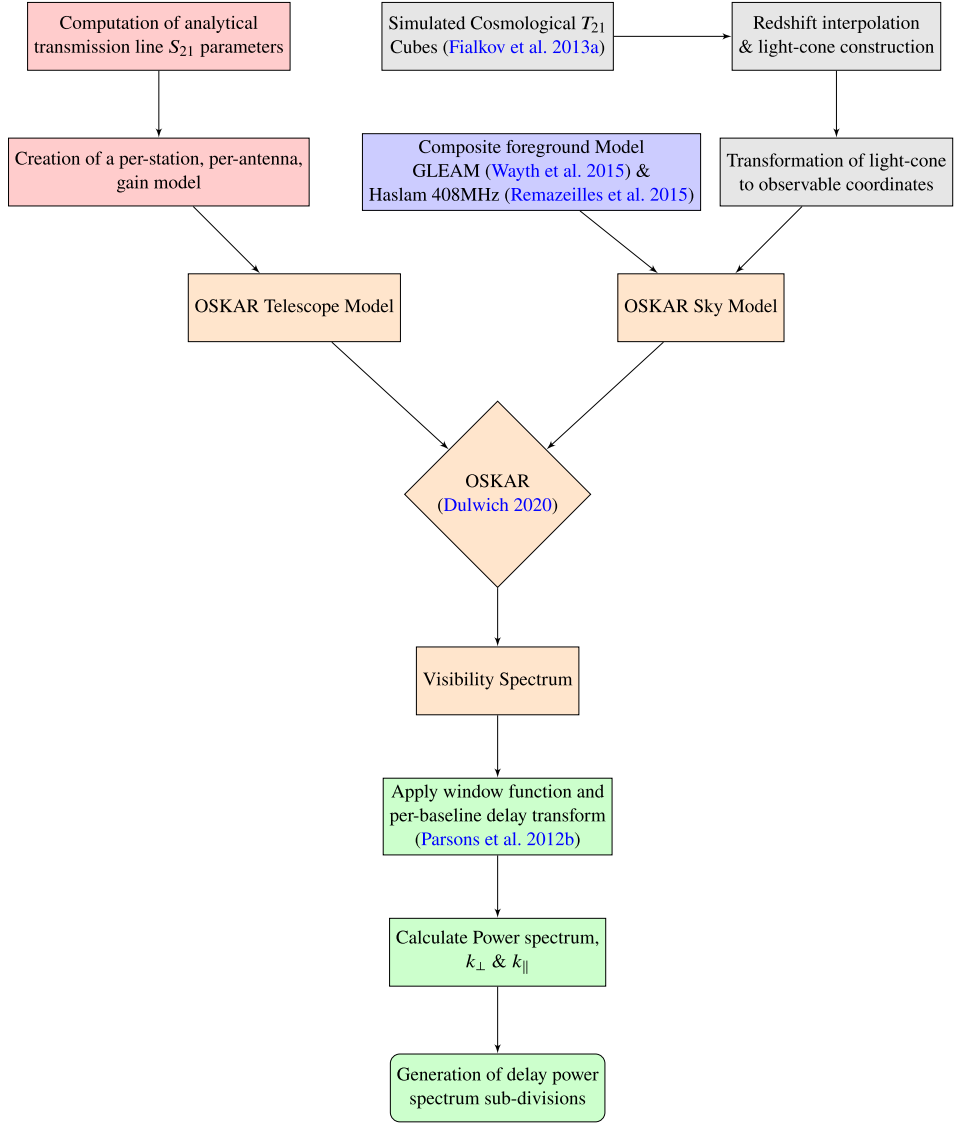

**Figure 2.** Project pipeline workflow: analytically computed complex cable gains (Section 2.4, in red), upgraded resolution sky maps from Haslam and GLEAM source catalogues (Section 2.3, in blue), interpolated semi-numerical 21-cm simulations (Section 2.2, in grey), OSKAR simulator, with its inputs and outputs (Section 2.1, in orange), and post-processing including a per-baseline delay transform and calculation of various summary statistics (Section 2.5, in green).

## 2 END-TO-END SIMULATION PIPELINE

The various stages of the end-to-end simulation pipeline workflow are presented in Fig. 2. In this section, we aim to outline in detail each of these stages. We first discuss the theory of operation of the OSKAR telescope simulator in Section 2.1. This includes highlighting the various inputs required to simulate visibilities for SKA-Low. Sections 2.2 and 2.3 explain the motivation and methodology behind the construction of the composite sky model, containing radio foregrounds from the Galactic and Extragalactic All-Sky MWA (GLEAM) Survey and Haslam 408 MHz, and a semi-numerically simulated 21-cm cosmological signal. In Section 2.4, we provide the formulae that we use to analytically compute systematic transmission line effects as a per-station complex gain model. Finally, Section 2.5 describes the theory behind the delay power spectrum $P_d$, while outlining a set of closed-form equations to calculate the interferometer response associated with a given $k_\perp$, $k_\parallel$ Fourier sky mode.

### 2.1 Simulating mock visibility data

We use the OSKAR software package (Dulwich 2020) to generate simulated visibilities, which are the primary inputs required for our EoR data processing and analysis pipeline. OSKAR is a GPU-accelerated simulator developed specifically to simulate observations made with large radio interferometers consisting of aperture arrays, where collections of elements are first digitally beam-formed before the station beam data are cross-correlated, as envisaged for the SKA-Low telescope. The Radio Interferometer Measurement Equation (RIME; Hamaker & Bregman 1996; Hamaker, Bregman & Sault 1996; Sault, Hamaker & Bregman 1996) provides the mathematical framework that allows us to model instrumental effects in a way that reflects their physical origins, and obtain the interferometric response using a sum over discrete sources across the whole sky. Visibility data are generated on all simulated baselines using the Jones matrix form of the RIME, as described, for example, in Smirnov (2011).

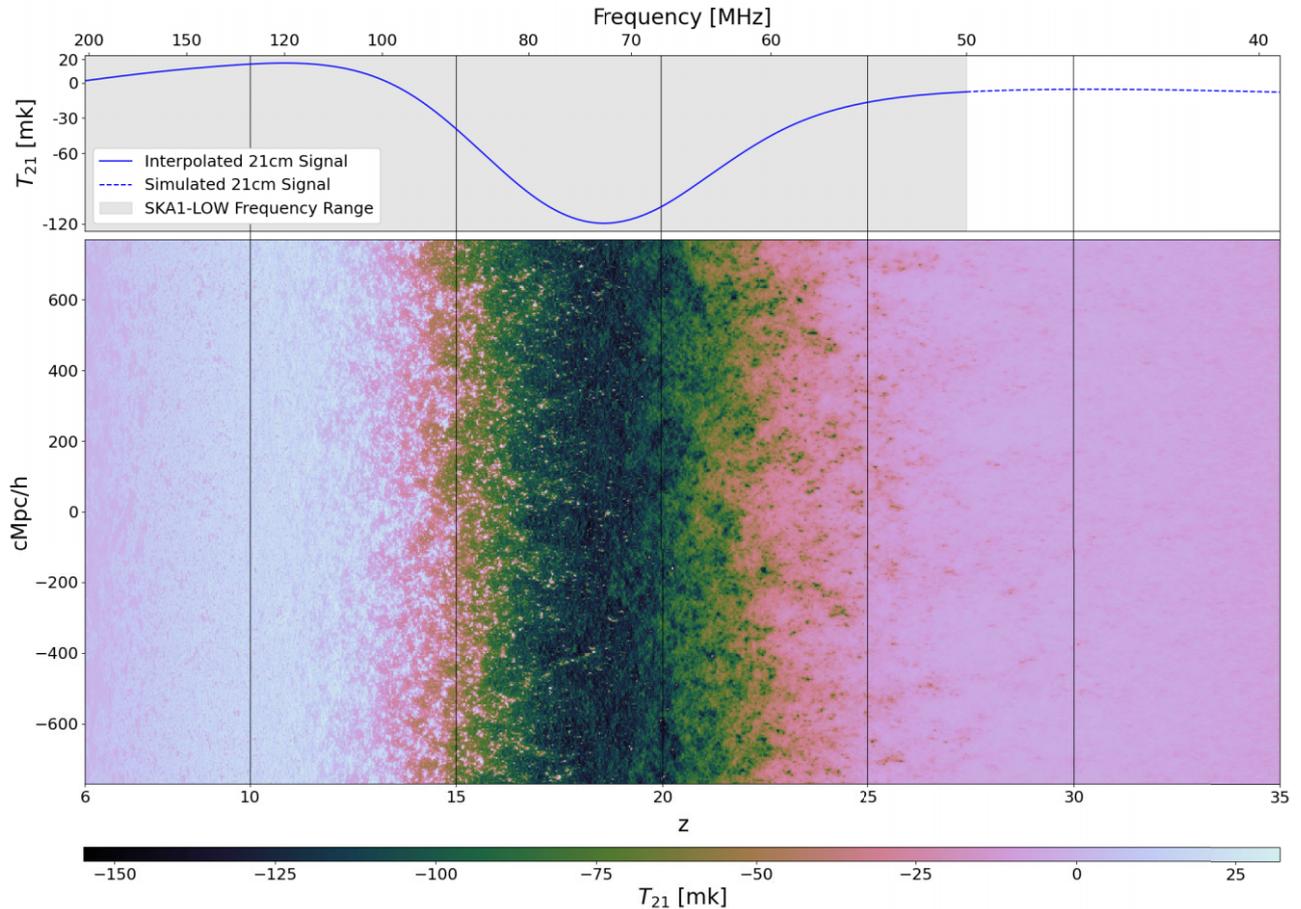

**Figure 3.** An illustration depicting the redshift evolution of the global and spatially varying 21-cm cosmological signal. The simulated data cubes of 21-cm brightness temperature (as discussed in Section 2.2) are generated from the Cosmic Dawn ($z = 50$–$15$) to the end of the EoR ($z = 15$–$6$), with a redshift resolution of 1 and 0.1 for the two regimes, respectively (Fialkov et al. 2013). Thus, to obtain 21-cm cubes with sufficient frequency resolution matching that of SKA-Low ($\Delta B = 109.8$ kHz; Trott & Wayth 2016), a per-pixel cubic-spline interpolation was applied. A light-cone was constructed using $512^3$ cubes to track the evolution of 21-cm brightness temperature along the line of sight, and a 2D slice is shown in the bottom panel. The global signal is then obtained by spatially averaging the resulting light-cone, and is displayed within the top panel. At a redshift of 6, the simulation box was fully ionized, resulting in a vanishing 21-cm signal. As a result, the End-to-End pipeline simulations are limited to frequencies between 50 and 202.86 MHz.

As inputs for its simulations, OSKAR requires a sky model and a telescope model. For the latter, we obtained the coordinates of the 224 core station centre positions from revision 3 of the SKA-Low layout (Dewdney & Braun 2016). Each station is modelled as a unique set of 256 antennas arranged randomly inside a 38-m diameter circle, with the minimum spacing between antennas set to 1.5 m (Mort et al. 2016). Although OSKAR is capable of using different patterns for each antenna, for these simulations, we model the antenna responses as a simple half-wavelength dipole in both the $X$ and $Y$ polarizations. Each station beam is then generated using a combination of the isolated element patterns and the array factor, which is obtained by multiplying the antenna responses by suitable complex phase weights. The corruptions introduced by the physical properties of the cables (discussed in Section 2.4) are modelled by supplying per-element complex gains as a function of time and frequency, which have the effect of distorting the station beams evaluated as part of the RIME. The form of the RIME used to generate the visibility $\mathbf{V}_{p,q}$ between stations $p$ and $q$ is thus the discrete sum over sources $s$ in the sky model,

$$V_{p,q} = \sum_s K_{p,s} E_{p,s} B_s E_{q,s}^{\mathrm{H}} K_{q,s}^{\mathrm{H}}, \qquad (2)$$

where the Jones term $K_{p,s}$ is the standard interferometric phase term, and $E_{p,s}$ is the station beam response (evaluated as described above), and where both are functions of station index $p$ or $q$ and source $s$. The source brightness is $B_s$, and superscript 'H' denotes Hermitian transpose.

In the work presented here, we assume that the effect of the cable reflections is the same in both the $X$ and $Y$ antenna polarizations, and therefore the Jones matrix terms in the RIME can be treated as complex scalars. The finite channel width is used when generating the simulated visibilities to model the effects of bandwidth smearing in each channel, where the smearing varies both across the field of view and as a function of baseline length. Following the analytical expressions for the bandwidth smearing given in Thompson, Moran & Swenson (2017), we assume that each channel has a flat band-pass.

## 2.2 21-cm signal simulations

As discussed in Section 1, the robust detection of the cosmological 21-cm power spectrum is a primary scientific goal of SKA-Low (Koopmans et al. 2015). Thus, to gauge the extent to which cable reflection systematics may hinder this goal, we incorporate a mock 21-cm signal in our simulation pipeline.

Neutral atomic hydrogen in the early Universe produces the cosmological 21-cm signal by absorbing or emitting photons at its 21-cm spectral line (Madau, Meiksin & Rees 1997). The resulting excess, or deficit, of 21-cm photons is then redshifted by the expansion of the Universe out of the spectral line, preventing these photons from interacting further with the neutral hydrogen, and allowing them to survive unimpeded until the present day. Furthermore, this redshifting has the added advantage of allowing different times in the early Universe to be probed through different frequencies of radio photons seen today. Due to the large abundance of elemental hydrogen in the Universe, the 21-cm signal is anticipated to be one of the most promising probes of the Universe between recombination and reionization.

The predicted strength of this 21-cm signal depends primarily on the abundance of neutral hydrogen, the occupancy of its upper and lower hyperfine states, and the relative temperature of the radio background. Hence, modelling the cosmological signal requires determining the temporal and spatial variations of these quantities, and thus consideration of the cosmological initial conditions of the Universe, radiative transfer, the atomic physics determining the hyperfine occupancy levels, and potential novel physics such as interactions between matter and dark matter (Barkana 2018). A detailed discussion of the theory of the cosmological 21-cm signal is beyond the scope of this paper (an interested reader may want to read the following review articles: Madau et al. 1997; Furlanetto et al. 2006; Pritchard & Loeb 2012; Barkana 2016; Mesinger 2019); all we require in this work is an example 21-cm signal cube to provide a reference for the strength of the anticipated cosmological signal.

We generate our example cosmological 21-cm signal cubes utilizing the semi-numerical code, 21cmSPACE (e.g. Visbal et al. 2012; Fialkov et al. 2014a; Reis et al. 2020) with the parameters outlined in Table 1. The code simulates a region of the early Universe by dividing it into cubic cells with side lengths of 3 comoving megaparsec (cMpc). For each cell, dark matter collapse into haloes and star formation is modelled by subgrid analytical prescriptions, while radiative transfer between cells is modelled numerically. This hybrid approach allows the simulation to probe cosmological volumes while including the small-scale physics that is key to simulating the 21-cm signal. An up-to-date description of the code and its features can be found in Gessey-Jones et al. (2023).

For the 21-cm signal simulation used throughout this paper, stars are assumed to form in haloes with critical circular velocities greater than $4.2 \, \mathrm{km \, s^{-1}}$. Population II and Population III stars are modelled separately with star formation efficiencies of 0.05 and 0.002, respectively, and the Population II star formation efficiency is log-suppressed in haloes below the atomic cooling threshold of $16.8 \, \mathrm{km \, s^{-1}}$ (Fialkov et al. 2013). The transition between these stellar populations is assumed to take 30 Myr due to supernovae ejecting the metal-enriched gas (Magg et al. 2022), and the Population III stars are modelled as having a log-flat initial mass function (IMF; Gessey-Jones et al. 2022). Furthermore, all galaxies in the simulation are assumed to emit X-rays at an efficiency of $3 \times 10^{40} \, \mathrm{erg \, s \, yr^{-1} \, M_\odot^{-1}}$ with a power-law spectral energy distribution (SED) of exponent $-1.5$ and a lower cut-off of 0.1 keV (Pacucci et al. 2014; Fialkov, Barkana & Visbal 2014b), and to be radio sources with 10 times the radio emissivity per star formation rate as present-day star-forming galaxies (Reis et al. 2020; Sikder et al. 2024). The ionizing efficiency parameter $\zeta$ of these same galaxies is set to 15 and we assume the maximum mean free path of the ionizing photons in the intergalactic medium is 50 cMpc. The simulations modelled Lyman heating (Reis, Fialkov & Barkana 2021), CMB heating (Venumadhav et al. 2018), and X-ray heating of the intergalactic medium while including photoheating feedback (Cohen, Fialkov & Barkana 2016), accounting for the multiple scattering of Ly$\alpha$ photons (Reis et al. 2021), modelling the relative velocities between dark matter and baryons (Visbal et al. 2012). The Lyman–Werner feedback is enabled (Fialkov et al. 2013; Muñoz et al. 2022).

To avoid tiling simulation boxes, and the excess power this introduces, we require the SKA-Low station beam to fit entirely within our 21-cm signal simulation boxes at all redshifts of interest. Hence, we use a simulation cube with $N_{\mathrm{pixel}} = 512^3$. Because each pixel has a side length of 3 cMpc, each side of the cube totalled 1.5 cGpc in length.[1] The typical size $128^3$ simulation box is commonly employed in forecasts and theoretical investigations (Reis et al. 2021; Gessey-Jones et al. 2023; Pochinda et al. 2024). The larger box has the additional benefit of giving us reliable statistics for the cosmological signal on larger scales. To cover the full frequency range of SKA-Low, the simulation was run from redshift 50 (27.8 MHz) down to 6 (203 MHz), outputting 21-cm signal cubes with a redshift resolution of 0.1 for values less than $z = 15$, but redshift values above this range are restricted to integers. At redshift 6, the Universe is assumed to be almost fully ionized, resulting in a near-vanishing 21-cm signal. While the latest observations disagree with this assumption evidencing a late end to reionization beyond redshift 6 (Bosman et al. 2022), the 21cmSPACE simulations are restricted beyond this point, and our investigations of cable reflections are limited to frequencies below 203 MHz. An interferometric array will observe a projection of the 21-cm signal evolution along the line of sight. The light-cone found in Fig. 3 is constructed from the 21-cm signal cubes, by applying a cubic interpolation along the line of sight to match the frequency range and resolution of SKA-Low power spectrum experiments ($\Delta B = 109.8$ kHz across the 50–350 MHz observation range; Trott & Wayth 2016). The light-cone is then converted to local observable coordinates before being combined with the foreground sky model.

We exclude the longest baselines to avoid probing angular sizes smaller than the pixel separation in the projected 21-cm signal. In our calculations, we assume that the arc formed by the subtended angle of the field of view ($\sim 5°$) on the sky may be approximated by the corresponding chord; for the SKA-Low primary beam, this approximation deviates on the order of $10^{-4}$.

## 2.3 Foreground model

To accurately approximate the foregrounds accessible to the SKA-Low, a sky model with sufficient angular and frequency resolution must be used.

Initially, we draw upon the work of the MWA, an SKA-Low precursor (Ewall-Wice et al. 2016), and their GLEAM survey (Wayth et al. 2015; Hurley-Walker et al. 2016), a catalogue of 307 455 radio sources spanning 72–231 MHz at $\approx 2$ arcmin resolution (Hurley-

---

[1]The simulations are largely memory-intensive and their computational requirement scales as $N_{\mathrm{pixel}}^3$. Hence, the standard $128^3$ sized simulation boxes take $\approx 3$ h using 20 GB RAM, while the $512^3$ boxes take $\sim 8$ d using 1 TB RAM, on $2\times$ AMD EPYC 7763 64-Core Processor 1.8 GHz.

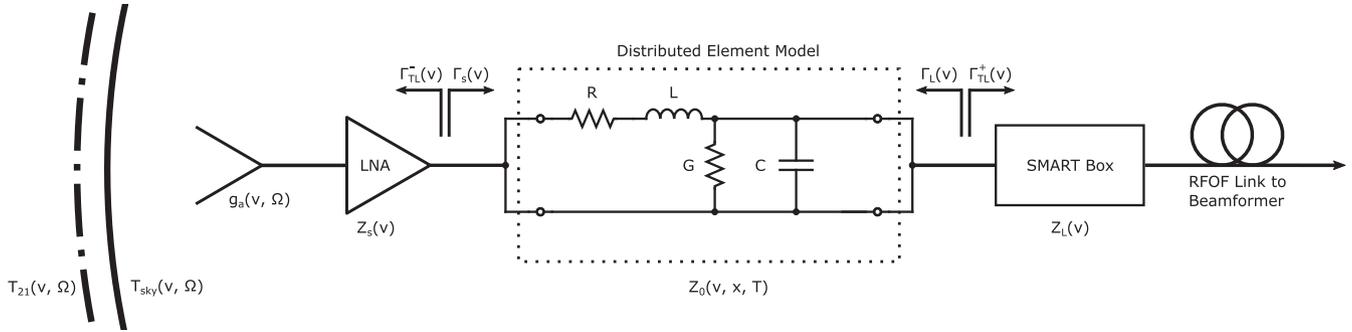

**Figure 4.** A circuit diagram describing the transmission line distributed element model of a single SKA-Low element. The antenna response $g_a$ on the sky consisting of a composite foreground model $T_{sky}$ and 21-cm cosmological signal $T_{21}$ is dependent on the scan angle $\Omega$ and the frequency of observation $\nu$. The resulting analogue signal propagates through the Low Noise Amplifier (LNA) and the coaxial transmission line before arriving at the SMART box. However, impedance mismatches between $Z_s$, $Z_0$, and $Z_L$ results in non-zero reflection coefficients $\gamma$ and chromatic cable reflections are observed in the instrument passband. The analogue ∼ 5 km RF-over-fibre (RFoF) link connects the field node to the signal processing subsystem for beamforming (Labate et al. 2022; Perini et al. 2022; Wayth et al. 2022).

Walker et al. 2016). In addition to the point sources from GLEAM, we need to include a diffuse component. The Global Sky Model (GSM; de Oliveira-Costa et al. 2008; Zheng et al. 2017) is a leading effort to model diffuse Galactic radio emission. It provides foreground data extending from 10 MHz to 5 THz, with 56 arcmin resolution below 10 GHz, by linearly interpolating a set of sky maps. Across the observational range of SKA-Low, two discontinuities are observed at 85 and 150 MHz in the interpolated spectral indices, which may impart additional frequency-dependent structure. Furthermore, the GSM has not been de-sourced, which would result in the double counting of sources that are also in GLEAM. For these reasons, instead of the GSM, we opt to use the de-sourced, de-striped Haslam 408-MHz map (Remazeilles et al. 2015), a spectral-index-based frequency-scaled model of diffuse Galactic synchrotron radiation. PyGDSM (Price 2016) was used to generate the scaled Haslam 408-MHz map with an included −2.725 K constant offset to account for the microwave background component.

We treat all sky model components and pixels as unresolved point sources. Therefore, to simulate a physical diffuse component for the ∼ 74-km maximum baseline of SKA-Low, a total of ∼ $2 \times 10^{11}$ sources (or pixels) would be required across the whole sky. For interferometric arrays, the shortest baselines provide access to the majority of the k-modes of the EoR window. Thus, to reduce the computational complexity of our sky model, we restrict the array configuration to the ∼ 1-km baselines in the core. The diffuse sky map is loaded into HEALPY (Zonca et al. 2019) and upscaled to $n_{side} = 2048$ (∼ $5 \times 10^8$ total sources) before being combined with the 21-cm light-cone and the GLEAM catalogue inside OSKAR.

### 2.4 Analytical coaxial transmission line model

Analysing the network response of a system front-end to account for cable reflections (Shi, Tröltzsch & Kanoun 2011) and the attenuation effect (Ellingson et al. 2018; Ellingson 2020) typically requires the use of external EM solvers such as CST (CST Microwave Studio 2019), COMSOL,[2] and Simulink.[3] To achieve independence of the end-to-end pipeline from these external solvers, we construct an analytical distributed element transmission line model, as shown in Fig. 4, to calculate the observed analogue signal at the beamformer from the composite material bulk properties and line geometries.

Coaxial cables are wide-band, single-ended, high-isolation transmission lines consisting of an inner cylindrical and outer tubular conductor with radii $\mathcal{A}$ and $\mathcal{C}$, respectively, carrying transverse electromagnetic (TEM) modes. The conductors are separated by a continuous low-loss dielectric of radius $\mathcal{B}$, enclosed by an insulating outer jacket (Kliros 2011). The outer conductor behaves as a continuous Faraday shield, attenuating external EM radiation with wavelengths shorter than the material skin depth. This isolates the core from environmental radio frequency interference (RFI; Xiao, Du & Zhang 2019), making it an ideal cost-effective solution for bounded RF transmission.

The telegrapher's second-order coupled differential equations describe the frequency-domain behaviour of current $\tilde{I}$ and voltage $\tilde{V}$ in a transmission line with respect to the EM properties of distributed-element equivalent circuits (Pozar 2011; Balanis 2012; Collin 2007)

$$\frac{d\tilde{V}(x)}{dz} = -[R + j\omega L]\tilde{I}(x_l), \quad (3)$$

$$\frac{d\tilde{I}(x)}{dz} = -[G + j\omega C]\tilde{V}(x_l). \quad (4)$$

Here, the angular frequency is $\omega = 2\pi\nu$ and the propagation distance from the load is denoted by $x_l$. The phasors are analogous to the dispersion-less wave equation whose propagation constant $\gamma$ is composed of real and imaginary components representing the attenuation $\alpha$ and phase $\beta$ constants

$$\gamma = \alpha + j\beta = \sqrt{(R + j\omega L)(G + j\omega C)}. \quad (5)$$

The series resistance $R$, the series inductance $L$, the shunt conductance $G$, and the shunt capacitance $C$ can be represented using closed-form equations given the cable structure and composite material bulk parameters.

Analytically, the transmission line theory of coaxial cables can be characterized by two frequency-dependent distinctive regimes dominated by the series resistance $R$ and inductance $L$ (Ramo, Whinnery & Van Duzer 1994; Grover 2004). At high frequencies ($\nu > 1$ MHz), the skin depth $\delta = \sqrt{2/\sigma\omega\mu}$ is negligible compared to the radial thickness of either conductor described by the conductivity $\sigma$ and permeability $\mu$. The current is confined to internal surfaces in contact with the dielectric, minimizing the inductance but maximizing the path resistance. At low frequency ($\nu < 20$ kHz), however, the opposite is true: the comparatively large skin depth allows the current to flow through the entirety of the conductors. Schelkunoff (1934) provides a comprehensive description of modelling both regimes and transition regions by employing Bessel functions to solve for the magnetomotive intensity. However, this approach is computationally taxing, and with the frequency range of SKA-Low extending from 50 to 350 MHz, this full analytical solution may be disregarded in favour of the high-frequency approximation (Ramo et al. 1994)

$$R = \frac{1}{2\pi} \left[ \frac{\rho_{\rm in}(1 + \xi \Delta T)}{\delta_{\rm in} \mathcal{A}} + \frac{\rho_{\rm out}(1 + \xi \Delta T)}{\delta_{\rm out} \mathcal{B}} \right], \quad (6)$$

where $\rho$ is the conductor resistivity measured at a reference temperature $T_{\rm ref}$, $\xi$ is the temperature coefficient of resistance, and $\Delta T = T - T_{\rm ref}$ is the temperature deviation from the reference temperature. Thus, any potential thermal effects within the cables can be accounted for, such as the diurnal variation of the ambient air temperature (e.g. at the Murchison Radio-astronomy Observatory, this varies between $-10°C$ and $+50°C$; Perini et al. 2022) or the local temperature change due to the long baselines extending upwards of 65 km (Labate et al. 2017).

---
[2]See https://www.comsol.com/comsol-multiphysics.
[3]See https://www.mathworks.com/products/simulink.html.

The inductance may be subdivided into two components, originating from the magnetic flux density inside (internal inductance $L_{\rm i}$) and between (external inductance $L_{\rm e}$) the two conductors (Ramo et al. 1994; Grover 2004):

$$L = \underbrace{\frac{\mu_0}{2\pi} \ln\left(\frac{\mathcal{B}}{\mathcal{A}}\right)}_{L_{\rm e}} + \underbrace{\frac{\rho}{2\delta\pi\omega} \left(\frac{1}{\mathcal{B}} + \frac{1}{\mathcal{A}}\right)}_{L_{\rm i}}. \quad (7)$$

At high frequencies, $L_i$ tends towards zero and may be disregarded, as the propagating current is bound to the conductor surface, resulting in a constant inductance per unit length.

The frequency-dependent complex permittivity $\hat{\epsilon}$ for lossy media is given by (Seybold 2005)

$$\hat{\epsilon}(\omega) = \epsilon'(\omega) + j\epsilon''(\omega) = \epsilon_r(\omega)\epsilon_0 + j\frac{\sigma(\omega)}{\omega}. \quad (8)$$

The real and imaginary components $\epsilon'$ and $\epsilon''$ are determined by the permittivity of free space $\epsilon_0$ and material relative permittivity $\epsilon_r$ and conductivity $\sigma$. The dielectric material inside the coaxial transmission line is usually selected to minimize dielectric conductivity and to confine the EM signal to the conductors. Across the spectral range of interest to SKA-Low, the insulating low-density polyethylene within an RG-58u coaxial cable exhibits a constant complex permittivity (Young 1989) and a conductivity of $\sigma = 10^{-13}$ (CST Microwave Studio 2019). This gives rise to a frequency-independent attenuative loss known as the dielectric conduction $G$, and is defined per unit length:

$$G = \frac{2\pi\sigma}{\ln(\mathcal{B}/\mathcal{A})}. \quad (9)$$

The shunt capacitance $C$ controls the attractive and repulsive coupling impact of electrons between conductors, and affects the wave group velocity and magnitude:

$$C = \frac{2\pi\epsilon}{\ln(\mathcal{B}/\mathcal{A})}. \quad (10)$$

The nominal impedance for RF systems is typically confined to 50$\Omega$, the approximate geometric mean between minimizing the impedance-dependent attenuation ($Z_0 = 77\Omega$) whilst maximizing peak power handling ($Z_0 = 30\Omega$) (Guile & Paterson 1969). It is defined by taking the ratio of the coupled voltage and current phasors outlined in equations (3) and (4),

$$Z_0 = \sqrt{\frac{R + j\omega L}{G + j\omega C}}. \quad (11)$$

Commercial manufacturers aim to minimize the frequency-dependent deviation between characteristic impedance $Z_0$ and nominal impedance, reducing potential impedance mismatch effects. However, over wide-band operation this becomes implausible. For a lossy transmission line of length $x$ connected to a network analyser with port reference impedance $Z_{\rm ref}$, the scattering matrix (S-matrix) $S_{ij}$ may be constructed from the aforementioned closed-form equations (Friar 2000):

$$\begin{bmatrix} S_{11} & S_{12} \\ S_{21} & S_{22} \end{bmatrix} = \frac{1}{(Z_0^2 + Z_{\rm ref}^2)\sinh(\gamma x) + 2Z_0 Z_{\rm ref} \cosh(\gamma x)}$$
$$\times \begin{bmatrix} (Z_0^2 - Z_{\rm ref}^2)\sinh(\gamma x) & 2Z_0 Z_{\rm ref} \\ 2Z_0 Z_{\rm ref} & (Z_0^2 - Z_{\rm ref}^2)\sinh(\gamma x) \end{bmatrix}. \quad (12)$$

Suppose $S^{(A)}$ and $S^{(B)}$ are scattering matrices corresponding to two-port components, which constitute a given analogue-receiving chain. By taking the Redheffer star product between two connected components, the network S-matrix can be obtained (Rumpf 2011):

$$\begin{aligned} S^{(AB)} &= S^{(A)} \star S^{(B)} \\ &= \begin{bmatrix} S_{11}^{(AB)} & S_{12}^{(AB)} \\ S_{21}^{(AB)} & S_{22}^{(AB)} \end{bmatrix} \\ &= \begin{bmatrix} S_{11}^{(A)} + \frac{S_{12}^{(A)} S_{11}^{(B)} S_{21}^{(A)}}{1 - S_{11}^{(B)} S_{22}^{(A)}} & \frac{S_{12}^{(A)} S_{12}^{(B)}}{1 - S_{11}^{(B)} S_{22}^{(A)}} \\ \frac{S_{21}^{(B)} S_{21}^{(A)}}{1 - S_{22}^{(A)} S_{11}^{(B)}} & S_{22}^{(B)} + \frac{S_{21}^{(B)} S_{22}^{(A)} S_{12}^{(B)}}{1 - S_{22}^{(A)} S_{11}^{(B)}} \end{bmatrix}. \end{aligned} \quad (13)$$

The scattering matrices and frequency response of the SKALA4.1 LNA and SMART receiver are classified for project use only. Therefore, the scattering parameters and complex linear gain,

$$\mathcal{G} = S_{21}, \quad (14)$$

shall be calculated henceforth using equation (12).

Given a reference impedance of 50 $\Omega$, the simulated and measured reflection coefficient for the SKALA4.1 antenna ports vary between $-7$ and $-25$ dB (Raghunathan et al. 2023) across the restricted 72–203 MHz simulation band. Implementing equation (1) yields a potential impedance mismatch of $16.63 - 0.32\,\Omega$. Fig. 5 illustrates the gain and delay spectra for line lengths 8, 10, and 15 m, given a realistic intermediary impedance mismatch of 5$\Omega$ across the frequency range of SKA-Low. These gain models are then supplied as part of the OSKAR telescope model to assess the impact of the spectral artefacts highlighted in the delay spectrum. Although cable degradation and cracking should be detectable by station quality assessment, and the majority of their effects should be mitigated by station beam calibration, simulating these processes would require major additions to this pipeline, and thus is left for future work.

### 2.5 Per-baseline delay and EoR window

For a point source $\hat{s} = \langle l, m, n \rangle$ located on the sky, the antenna elements $\boldsymbol{p} = \langle p_x, p_y, p_z \rangle$ and $\boldsymbol{q} = \langle q_x, q_y, q_z \rangle$ define an interferometric baseline

$$\boldsymbol{b} = \langle p_x - q_x,\ p_y - q_y,\ p_z - q_z \rangle. \quad (15)$$

The antenna elements may observe a relative delay $\tau_{\rm g}$ for a propagating signal due to a difference in displacement from the associated

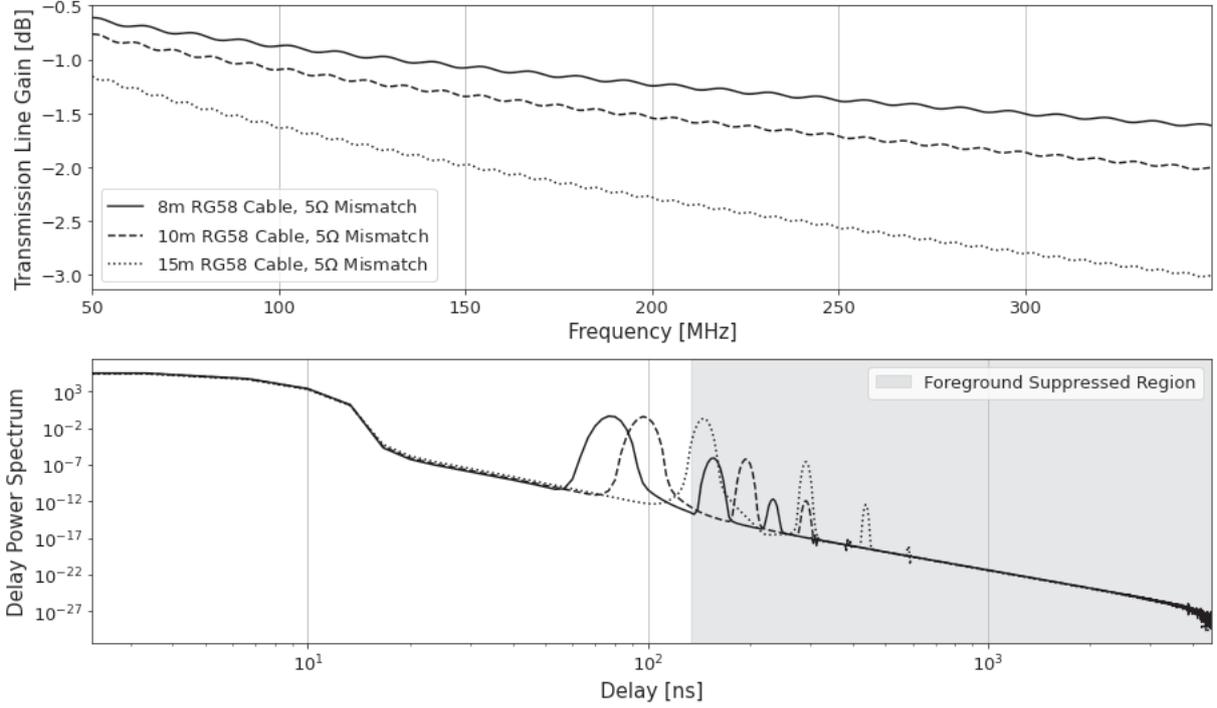

**Figure 5.** A simulated 8-m, 10-m, and 15-m RG-58u coaxial transmission line distributed element model spanning the 50–350 MHz observational window of SKA-Low. (a) A sinusoidal chromatic effect is introduced into the scalar linear gain resulting from cable reflections caused by a 5Ω impedance mismatch. (b) The delay power spectrum is obtained after tapering with a squared Blackman–Harris window function, and then taking the Fourier transform. The grey-shaded area represents the EoR window, which is defined by the horizon limit given for the minimum SKA-Low core station baseline $\boldsymbol{b}_{min} = 40.04$ m at redshift $z = 8$. The peaks in the power spectrum indicate the DC component and cable reflections up to the fourth order.

source,

$$\tau_g = \frac{\boldsymbol{b} \cdot \hat{\boldsymbol{s}}}{c} = \frac{1}{c}(b_x l + b_y m + b_z \sqrt{1-l^2-m^2}) = \frac{|\boldsymbol{b}|\sin(\theta)}{c}, \quad (16)$$

where $c$ is the speed of light and $\theta$ is the source zenith angle (Chapman et al. 2016). Equation (16) reveals the existence of a maximum geometric delay attributed to each baseline corresponding to a source located at the horizon, and is known as the 'horizon limit'.

Assuming $\hat{\boldsymbol{s}}$ is spectrally flat, the delay space $\tau$ signal response is described by a Dirac delta function $\delta_D(\tau_g - \tau)$. In reality, astrophysical emission spectra are not flat, and the interferometric response is inherently chromatic, broadening the delta function around the geometric group delay. This concept forms the basis by which the smooth foregrounds are separated from 21-cm emission (Parsons et al. 2012a, b; Liu & Shaw 2020). Beyond the horizon limit, spectrally smooth foreground power decreases rapidly, whereas the band-limited 21-cm emission can scatter power far beyond this geometric limit. An uncontaminated sample of 21-cm emission is obtained by observing power extending to delays beyond the horizon limit. Furthermore, the primary lobe of the interferometer beam is akin to an artificial horizon and is referred to as the 'beam limit'. The baseline delay transform $\widetilde{V}_b$ is computed for a given delay by taking the Fourier transform of the visibilities $V_{p,q}$ across the 20-MHz finite bandwidth frequency domain

$$\widetilde{V}_b(\tau) = \int W(\nu) V_{p,q}(\nu) e^{2i\pi\nu\tau} d\nu. \quad (17)$$

The window function $W(\nu)$ is a digital filter that tapers additional spectral components originating from discontinuities found at the boundaries of the finite dataset when calculating the Fourier transform. Furthermore, Lanman et al. (2020) show that a rectangular, Blackman–Harris (Harris 1978), and 120-dB Dolph–Chebyshev (Lynch 1997) window function do not possess the dynamic range required to suppress the foregrounds and the side-lobes of their Fourier dual below that of a fiducial 21-cm signal. Drawing from the power spectrum analysis implemented by Thyagarajan et al. (2016) and the HERA collaboration (DeBoer et al. 2017), we implement a modified Blackman–Harris window, convolving the window function with itself, $W(\nu) = W_{BH}(\nu) * W_{BH}(\nu)$. An specific example of delay-space structure of $W(\nu)$ is illustrated in Fig. 6.

To analyse the Fourier modes obtained from the sky, we differentiate the modes perpendicular ($k_\perp$) and parallel ($k_\parallel$) to the line of sight. The $k_\perp$ wavenumbers are determined by the angular scales accessible in the sky for a given distribution of baseline lengths, and the $k_\parallel$ modes correspond to the interferometer spectral information (Liu & Shaw 2020):

$$k_\perp = 2\pi \frac{\boldsymbol{b}\nu}{cD_c}; \quad (18)$$

$$k_\parallel = 2\pi \frac{H_0 E(z) \nu_{21} \tau}{c(1+z)^2}. \quad (19)$$

Here, the frequency of the 21-cm hydrogen line is given by $\nu_{21}$, and the Hubble parameter is represented as the Hubble constant $H_0$ times the dimensionless Hubble parameter $E(z) = \sqrt{\Omega_r(1+z)^4 + \Omega_m(1+z)^3 + \Omega_\Lambda}$. The comoving distance $D_c$ acts as a conversion factor to express the observational coordinates $\Omega$ and $\nu$ to the cosmological k-modes (Hogg 2000),

$$D_c \equiv \frac{c}{H_0} \int_0^z \frac{dz'}{E(z')}. \quad (20)$$

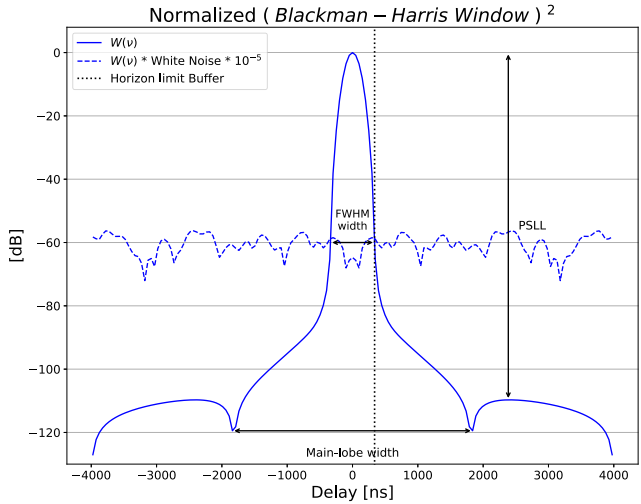

**Figure 6.** Illustration of the delay-space characteristics associated with a Blackman–Harris squared window function $W(\nu)$ for 160 channels each of width 125 kHz. The peak side-lobe level (PSLL), full width at half-maximum (FWHM), FWHM width, and main-lobe width were found to be $-109.71$ dB, $-60$ dB, 675.99 ns, and 3677.02 ns, respectively. The 337.99-ns primary lobe spillover, and thus the horizon limit buffer, are given by the primary lobe width when suppressed below the level of the proxy 21-cm signal (a magnitude-scaled white noise). For the case of $W(\nu)$, this suppression level corresponded with that of the FWHM.

Taking equation (16) and substituting $\tau$ and $\boldsymbol{b}$ in terms of their corresponding Fourier sky modes given by equations (18) and (19) for $\tau$ and $\boldsymbol{b}$, we obtain

$$k_\parallel = k_\perp \frac{D_c H_0 E(z)}{c(1+z)} \sin(\theta). \quad (21)$$

The linear proportionality between the parallel and perpendicular Fourier modes defines the gradient associated with the horizon and beam for a given depression angle $\theta$ (Parsons et al. 2012a; Chapman et al. 2016). These proportionality curves establish the range of delay for smooth foreground emission based on the geometric delay.

The delay power spectrum $P_d$ is obtained by calculating the energy spectral density of the natural visibilities, before scaling to the cosmological coordinate system for a given $k_\perp$ and $k_\parallel$,

$$P_d(k_\perp, k_\parallel) = |\widetilde{V}_b(\tau)|^2 \frac{A_e}{\lambda^2 \Delta B} \frac{D_c^2 \Delta D_c}{\Delta B} \left(\frac{\lambda^2}{2k_B}\right)^2. \quad (22)$$

Here, the antenna's fundamental properties, such as the antenna's effective area $A_e$, the channel bandwidth $\Delta B$, the band central wavelength $\lambda$, and the co-moving distance $\Delta D$, change for each band (Thyagarajan et al. 2015).

The key to sensitive measurements is to first combine samples of the same Fourier mode coherently, before adding the independent Fourier modes in quadrature (Parsons et al. 2012a). The SKA array configuration is designed to provide a smooth logarithmic distribution of baselines (Dewdney et al. 2013). Therefore, the visibilities are subsequently binned logarithmically across $k_\perp$, reducing the prevalence of vertical streaking within the EoR window associated with regions of poor UV coverage, as seen in Byrne et al. (2019). The resulting EoR window remains a promising region for the pursuit of the hyper-fine transition signature of hydrogen. The EoR window is, however, subject to chromatic instrumental effects such as cable reflections (Ewall-Wice et al. 2016) and mutual coupling (Josaitis

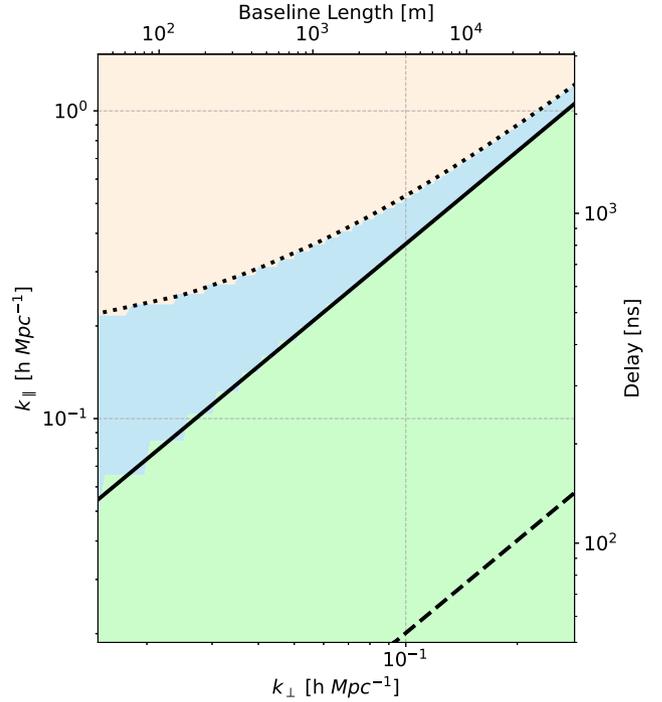

**Figure 7.** A graphical representation of the various k-mode regions of the delay power spectrum outlined within Section 3. The EoR window (shaded in a peach colour above the dotted line) is relatively devoid of smooth spectral foreground contamination, and contains the Fourier modes of the sky in which a detection of the 21-cm signal is most likely to occur. Thus, we choose to focus on the detectability of the 21-cm signal in the presence of potential foregrounds, and concentrate on the signal-to-foreground ratio. In the supra-horizon region (blue, below the dotted line), detectability is not our primary concern, but we are interested in understanding how the systematics affect the total power, and therefore we plot the absolute fractional difference between the power with systematics and the power without. The foreground wedge (green, lower shaded area below the sold line), the k-mode region found below the horizon limit, and the band-limited primary lobe of the window function are contaminated by smooth radio foregrounds. The horizon limit (black solid line), the constant 337.99-ns horizon limit buffer (black dotted line), and the beam limit (black dashed line) are calculated using equation (21) given the central frequency of the observation band $B$.

et al. 2022), which obscure the 21-cm signal and reduce the usability of the window.

## 3 REGIONS OF THE DELAY POWER SPECTRUM

The inclusion of a transmission line in the simulated system front-end results in a series of spectral artefacts, which contaminate the accessible k-modes of the delay power spectrum. The dielectric and resistive propagation losses within the transmission line attenuate the global power, while transmission line reflections push the power from one $k_\parallel$ mode to another. To examine the impact of using a non-ideal transmission line, we divide the delay power spectrum into three k-mode regions, as illustrated in Fig. 7. In the EoR window, our main focus is on the detectability of the signal in the presence of potential foregrounds, so we concentrate on the signal-to-foreground ratio. In the supra-horizon region, detectability is not our primary concern, but we are interested in understanding how the systematics affect the total power. Therefore, we plot the absolute fractional difference between the power with and without systematics for this region.

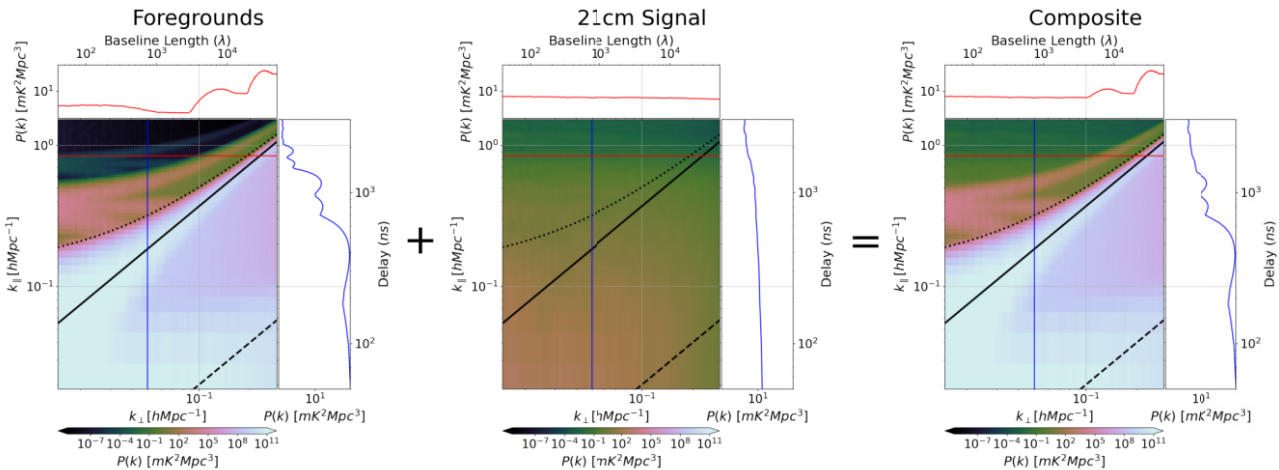

**Figure 8.** The End-to-End pipeline was used to simulate the delay power spectra for a pair of ideal instantaneous snapshots, corresponding to 160 channels across a 145–165 MHz observation range, using the SKA-Low core pointed at the MWA EOR1 field. The leftmost panel contains the composite foreground model described in Section 2.3, while the central panel contains the simulated 21-cm cosmological signal from Section 2.2. Limb-brightening along the horizon limit originates from an increase in foreground flux due to the Galactic Centre when it is far from the zenith. The rightmost delay power spectrum is a combination of both the foreground and 21-cm signals. The ability to straightforwardly combine these signals allows for statistical post-processing (outlined in Section 7) to be calculated with the aim of quantifying the effects of contamination outlined in Table 2. Cross-sections of constant $k_\parallel$ (red) and $k_\perp$ (blue) highlight the bounding effect of the beam (black-dashed) and horizon (solid-black) limits and the spectrally flat signature of the 21-cm cosmological signal. For delays that extend beyond 2.8 μs, the spectral structure of the 21-cm signal is no longer physical and would be dominated by the redshift-interpolation of the light-cone. This is shown in the middle panel by the signal along the $k_\perp$ slice fluctuating slightly at high $k_\parallel$ values.

### 3.1 The EoR Window

The k-modes above the horizon limit should exhibit the greatest ratio between the 21-cm signal and the radio foregrounds. Thus, we define a region of k-space, called the EoR window, in which for an ideal observation (i.e. without cable reflections; see Fig. 8) the signal-to-foreground ratio $\geq 5$ (Rose 2013). When analysing this region of k-space, we consider the effects caused by systematics at least $5\sigma$ above the 21-cm signal. The colour-bar lower bound of Figs 9(a), 10(a), and 11(a) are saturated at this criterion to highlight any apparent variation introduced. The redistribution of foreground power resulting from chromatic instrument systematics (e.g. cable reflections) should ideally be avoided during instrument design, to prevent the EoR window contamination and to maintain the highest probability of a successful 21-cm detection under foreground avoidance paradigms.

### 3.2 Supra-horizon dominated k-modes

Within the simulated delay power spectrum shown in Fig. 8, we observe an excess of foreground power that extends beyond the geometric horizon limit. The origin of this contamination, known as the supra-horizon structure, is poorly understood because of the combination of several chromatic effects within this region (Lanman et al. 2020), including the spectral behaviour of the beam and sky (Parsons et al. 2012b; Kern et al. 2020), the non-smooth spectral nature of realistic foregrounds (Pober et al. 2013; Jensen et al. 2016), and the spillover of the primary lobe of the window function and its 'effective' bandwidth (Chapman et al. 2016; Thyagarajan et al. 2016). In this work, the length of the simple half-wavelength dipole and the flux of the RF foreground emission are scaled according to the observation frequency, resulting in a spectrally smooth beam and sky response. The observed supra-horizon emission must therefore come from the implementation of the window function $W(\nu)$. Fig. 6 illustrates a 337.99-ns primary lobe spillover of $W(\nu)$, resulting from the 160 channels, each of width 125 kHz. The spillover width of the primary lobe was determined when the smooth DC components were suppressed below the level of the 21-cm signal. To isolate the k-modes that are free from foreground contamination from those dominated by supra-horizon emission, we follow the work of Barry et al. (2019) and Beardsley et al. (2016), and implement an additive horizon limit buffer according to this spillover.

The SKA has indicated its intention to focus on a foreground removal paradigm. To quantify the variation in power, the per-pixel absolute fractional difference $|P_{\mathrm{sys}} - P_{\mathrm{ideal}}|/P_{\mathrm{ideal}}$ was calculated. Here, $P_{\mathrm{sys}}$ and $P_{\mathrm{ideal}}$ correspond to an observation with an included systematic and an ideal uncontaminated observation, respectively. This allows us to evaluate whether calibration techniques such as point source removal, principal component analysis, and Gaussian process regression are necessary to account for attenuating transmission line effects, to avoid under-fitting or over-fitting of foreground data.

### 3.3 The foreground wedge

The k-modes below the horizon limit and within the band-limited primary lobe of the window function lie in the foreground wedge. Due to the extremely low signal-to-foreground ratio, we focus our analysis on the aforementioned regions of the delay power spectrum.

## 4 TRANSMISSION LINE FOREGROUND CONTAMINATION

For SKA-Low, the front-end component specifications and responses are classified for project use only, thus encouraging a parametric approach towards the transmission line modelling. This approach aims to organize and assess the structure of the foreground contamination beyond the geometric delay for various cable specifications.

We chose to implement a base cable model built from an 8-m RG-58u coaxial transmission line, which consists of a copper core (of ra-dius $\mathcal{A} = 0.4572$ mm), a low-density polyethene dielectric insulator (of radius $\mathcal{B} = 1.4732$ mm), and a braided copper screen/shield (of radius $\mathcal{C} = 1.7272$ mm) (values from CST Microwave Studio 2019). In Fig. 5(a), a cable-length-dependent chromatic oscillation can be observed in the scalar gain, while the peaks in the Fourier dual represent the DC component and the subsequent cable reflections towards the fourth order. To validate the accuracy of the analytical approach outlined in Section 2.4, numerical simulations were performed in CST Microwave Studio (2019) Cable Studio using a predefined RG-58u coaxial transmission line, and a reference impedance of 55 Ω. The numerical simulation was then compared to the analytical Python output, resulting in an RMSE < 5 per cent across the 50–350 MHz window, which deviated at the high frequencies because of the dielectric permittivity being sampled only at 100 MHz.

**Table 1.** A list of the astrophysical parameters utilized within this paper to model the 21-cm signal using 21cmSPACE as detailed in Section 2.2.

| Parameter | Value | Description |
|---|---|---|
| $V_c$ | 4.2 km s$^{-1}$ | Minimum virial circular velocity of a halo for star formation in the absence of feedback |
| $f_{*,II}$ | 0.05 | Pop II star formation efficiency |
| $f_{*,III}$ | 0.002 | Pop III star formation efficiency |
| $M_{atm}$ | 16.8 km s$^{-1}$ | Mass threshold at which star formation would occur in a halo due to cooling from atomic hydrogen |
| $t_{delay}$ | 30 Myr | Recovery time between Pop III and Pop II star formation |
| $f_X$ | $3 \times 10^{40}$ erg s$^{-1}$ M$_\odot^{-1}$ | X-ray emission efficiency of Pop II and Pop III star-forming haloes |
| $E_{min,II}$ | 0.1 keV | Pop II SED lower cut-off energy |
| $f_{rad}$ | 10 | Radio emission efficiency of high-redshift galaxies |
| $\zeta$ | 15 | Effective ionizing efficiency of high-redshift galaxies |
| $R_{max}$ | 50 cMpc | Effective ionizing efficiency of high-redshift galaxies |

**Table 2.** An overview of the parameters that define the standard simulated observation and their variations through the SKA End-to-End pipeline. The parameter selection aims to encompass potential transmission line contamination of the delay power spectra for SKA-Low, despite the computationally expensive foreground model (refer to Section 2.3) and frequency limitations of the simulated 21-cm signal (refer to Section 2.2). Figs 8, 9, 11, and 12 quantify the contaminating spectral artefacts of the k-modes within the EoR window, motivating future calibration and interferometric design.

| Standard observation | Values | Units |
|---|---|---|
| Array configuration | SKA-Low Core | |
| Observation time | 2000-01-01 13:31:30 UTC | |
| Right ascension | 60° | |
| Declination | −30° | |
| Observation range ($B$) | 145–165 | MHz |
| Channel bandwidth ($\Delta B$) | 125 | kHz |
| Cable type | RG-58u coaxial cable | |
| Cable length ($x$) | zero tolerance 8 metres | m |
| Impedance mismatch | 5 Ohms | Ω |
| Base temperature ($T_{ref}$) | Uniform 295.15 Kelvin | K |
| **Parameter variations** | | |
| Transmission line length ($x$) | [15, 25, 100, 1000] | m |
| Impedance mismatch | [0, 3, 10, 50] | Ω |
| 1$\sigma$ Length tolerance | [0.1, 1, 3, 5] | % |
| Thermal variation range ($T$) | [1, 3, 5] | ± K |

The visibilities and delay power spectra were simulated for all 24 976 instantaneous baseline pairs from the 224 core stations, using 160 channels (each of width 125 kHz) for frequencies between 145 and 165 MHz, at 2000-01-01 13:31:30 UTC. The phase centre, at RA = 60°, Dec. = −30°, corresponds to the direction of the MWA EOR1 field (Beardsley et al. 2016), and the field transits at the chosen observation time. This field was chosen as it is relatively devoid of Galactic emission and bright extragalactic sources, thus minimizing the foregrounds.

Once the standard cable model and observation were defined, the transmission line parameter space was varied to cover the range of spectral artefacts SKA-Low can observe. This simulated parameter space is outlined in Table 2, while the following subsections provide a statistical comparison of the delay power spectra following the k-space divisions discussed in Section 3.

### 4.1 Transmission line length and impedance mismatch

A perfect impedance match of lossless electrical components is a theoretical assumption. In reality, the propagating EM signal experiences an unavoidable loss and partial reflection at the transmission line port interface. The spectral response of the instrument is contaminated by these systematic length-dependent transmission line effects, as shown in Section 2.4. The RG-58u coaxial cable contains low-density polyethylene (LDPE), with a dielectric constant of 2.12 (value from CST Microwave Studio 2019). Therefore, the reflected EM signal has a group velocity of 2.06 ms$^{-1}$ given by $v_g \approx c/\sqrt{\epsilon_r \mu_r}$. For an 8-m coaxial cable, the characteristic impedance at 200 MHz was calculated to be 48.2 Ω, assuming a constant excess 5 Ω impedance mismatch at both ends (as shown in Fig. 5); the first reflection has a magnitude, relative to the initial signal, of $2.4314 \times 10^{-3}$ at a delay of 77.65 ns, and a second reflection of $5.9117 \times 10^{-6}$ at 155.31 ns. Thus, the first-order reflection is large enough to mask the 21-cm signal, while the second-order reflection introduces an observable variation in the signal-to-foreground ratio.

The length-dependent contamination of the delay power spectrum is demonstrated in Fig. 9, for transmission lines of lengths 8, 15, 25, and 100 m. We observe no apparent change within the EoR window for the 8-m line due to the confinement of relevant reflections to the foreground wedge. Although the delay peaks are buried below the horizon limit, line attenuation alters the foreground power as demonstrated in Fig. 9(b) and deviates from existing 'known' foreground calibration models. The reflection peaks attributed to the 15, 25, and 100 m lines move towards higher delays and extend beyond the horizon limit buffer. For the 100-m coaxial line, the contaminating foreground power masks a large proportion of the accessible k-modes composing the EoR window. The pitchfork structure associated with each reflection originates from the inherent chromaticity of the interferometer and is cross-validated with that of uncalibrated reflections within MWA for redshifts 7, 12, and 16 (Ewall-Wice et al. 2016).

Transmission lines whose first-order reflection exceeds the accessible delay space of the band-limited observation are subject to aliasing artefacts. For the 1-km lossless transmission line shown in Fig. 10, we demonstrate the impact of this aliasing, where power originating below the horizon limit buffer contaminates the EoR window. CST simulations calculate the attenuation of the base cable model to be −14.35 dB/100 m at 150 MHz (the band central frequency). Assuming that the 1-km cable is lossy, the first-order reflected signal experiences −430.5 dB of attenuation, and thus the resulting spectral artefact is no longer visible above the 21-cm signal, as shown in Fig. 10 (right panels). This effective filtering of the reflected signal may be achieved by using a ferrite attenuation bead for shorter transmission line lengths. This passive device filters high-frequency noise across a broad frequency range. Assuming that the resistive region of the bead spans the observational frequency band, the bead acts as a resistor, dissipating the propagating signal energy as heat.

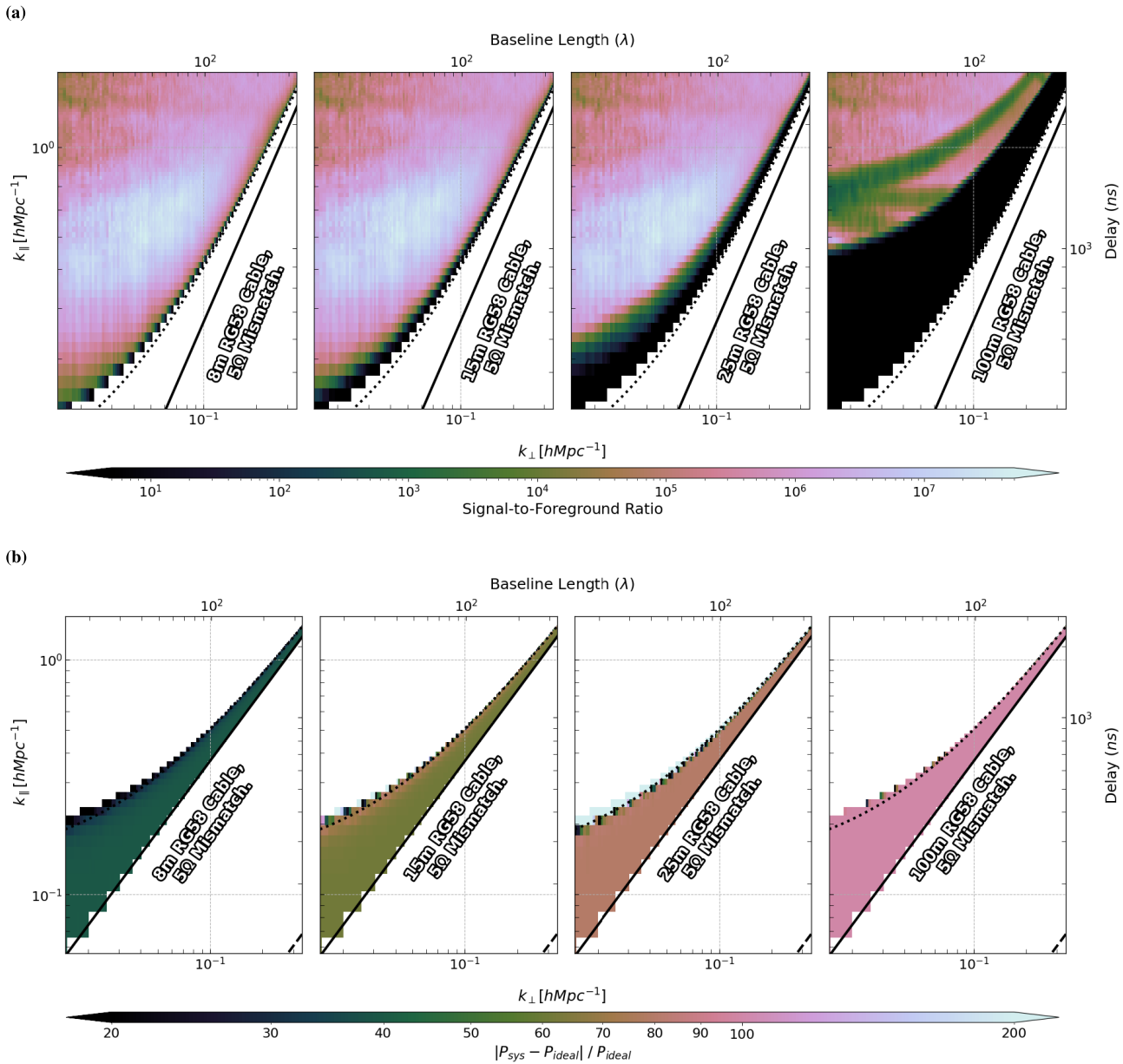

**Figure 9.** A parametric sweep of the transmission line length, as outlined in Table 2, illustrating the spectral contamination of (a) the EoR window and (b) the supra-horizon dominated k-modes. Along the top row, the signal-to-foreground minima are a manifestation of the delay peaks observed in Fig. 5, originating due to an impedance mismatch at the transmission line interface. The reflected power significantly reduces the signal-to-foreground ratio within the EoR window, contaminating the detectable 21-cm bins. The increasing transmission line length lifts these delay peaks, concentrated along the buffer line, towards higher delays/$k_\parallel$ modes, revealing their pitchfork structure. Transmission lines of length ≤ 15 m confine the reflection below the horizon limit buffer (dotted black line). However, along the bottom row, the absolute fractional difference in the k-modes deviates from 'known' calibration models.

Fig. 11 illustrates a parameter sweep of impedance mismatch for 0Ω, 3Ω, 5Ω, 10Ω, and 50Ω. The peak attributed to the first-order reflection for an 8-m transmission line cannot be observed directly and is buried within the band-limited foregrounds. The proportional increase in impedance mismatch, and thus reflected power magnitude, may lift higher-order reflections beyond the 21-cm signal. Due to the finite bandwidth of observation, and the discrete nature of the delay transform, the reflection peaks are broad. Therefore, given a sufficiently large impedance mismatch, the tails of the reflection peaks extend outwards, contaminating $k_\parallel$ modes surrounding the fundamental delay. An example of this behaviour can be seen in the rightmost panel of Fig. 11(a), where the tail of the second-order reflection centred at 155.31 ns extends beyond the horizon limit buffer. The SKALA4.1 antenna exhibits ≤ 16.6Ω mismatch across the selected simulation band; however, extreme impedance mismatches (≥ 50Ω) may occur in the system front-end due to transmission line cracking and degradation. While this work aims to quantify the impact of such adverse effects, we anticipate that they will be identified during quality control in construction and quality assessment metrics in station calibration. Consequently, these effects should not propagate through to the delay power spectrum. Even for the case of the perfectly matched front-end network shown in the leftmost panel of Fig. 11(b), an unavoidable ∼37 per cent absolute fractional difference is observed due to line attenuation, emphasizing the need for the inclusion of transmission line effects in the calibration model.

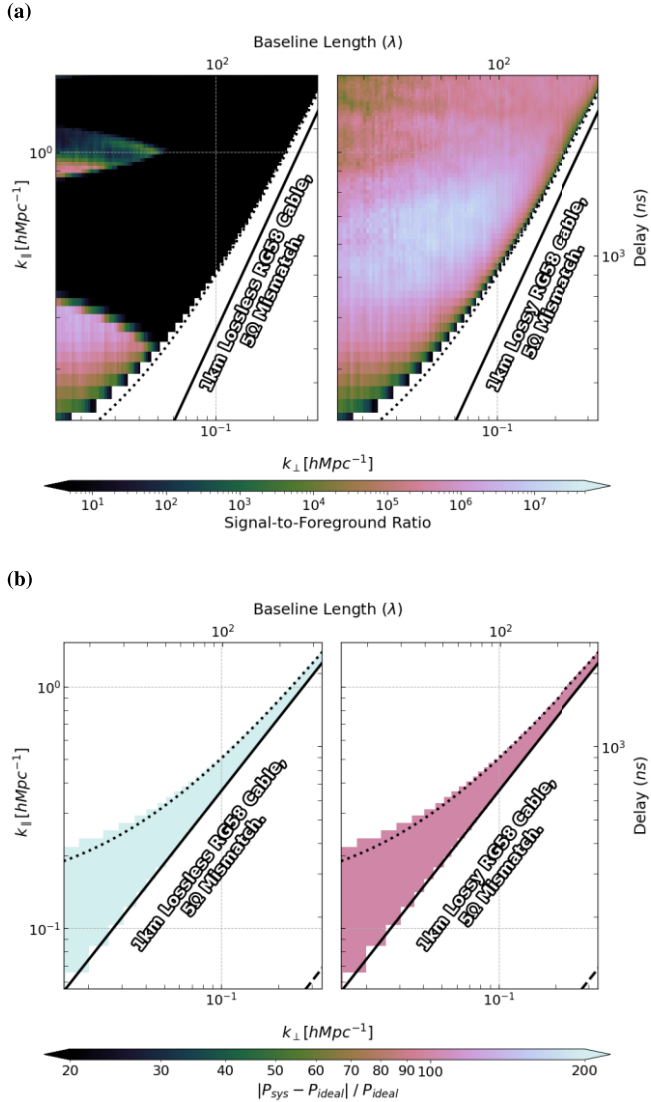

**Figure 10.** The contamination impact of aliasing across (a) the EoR window and (b) the supra-horizon dominated k-modes for a lossy and lossless 1-km transmission line, whose delay extends beyond the maximum bandlimited $k_\parallel$ mode. The resistive and dispersive attenuation found in the lossy transmission line reduces the amplitude of the reflected power below that of the 21-cm signal.

### 4.2 Manufacturing tolerance and thermal variation

The geographical scale of SKA-Low will result in environmental factors such as temperature, RFI, and the ionosphere varying across the instrument. While most of these factors do not directly affect the transmission line modelling (and are out of the scope of this paper), equation (6) exhibits a clear thermal dependency. Temperatures were sampled from a Perlin noise map (Perlin 1985; Ebert et al. 2003) with a 10 m$^2$ spatial resolution. The amplitude range of the noise map is symmetric and scaled across a variety of temperature ranges, ±[0, 1, 3, 5] K (see Table 2). Fig. 12(b) illustrates the absolute fractional difference between a reference observation, and an observation with thermal effects included for each case. Although a brightening along the horizon limit buffer is observed, this artefact arises from the absolute fractional difference calculation, where the window function varies fastest and the ratio of the values is large.

Variations due to manufacturing tolerances in front-end components are inevitable, which will further complicate the calibration of the station beam. For flexible and semi-rigid cable assemblies, the industrial manufacturing standards quote maximum coaxial transmission line length tolerances of 1.0 per cent, with tighter tolerances available at increased cost. To understand the effect on the station beam and thus the EoR window, we randomly sample the antenna cable length tolerance from a Gaussian distribution with widths shown in Table 2. The transmission line scattering parameters are calculated using equation (12), and this per-station, per-antenna gain model is provided to OSKAR. For each observation, the absolute fractional difference was calculated with respect to the reference observation outlined in Table 2 and shown in Fig. 12(a).

Across the accessible $k_\perp$ modes of Fig. 12(b) we observe structure in the differences of the power spectra. This structured noise-like component is superimposed on the smooth artefact in Fig. 12(a), where the magnitude of the noise-like component increases as the length tolerance becomes less stringent. We attribute this effect to each element possessing a unique gain, thus modifying the structure of each station beam and, in particular, the associated side-lobes. If a per-element gain correction could be used, this effect could be allowed for the calibration process. The cable reflections due to variations in length caused by manufacturing tolerance will also broaden the delay peak structure.

## 5 CONCLUSIONS AND FUTURE WORK

This paper details an End-to-End simulation pipeline designed to measure the instrumental effects of SKA-Low, whilst accounting for the frequency- and temperature-dependent cable parameters that may affect its performance. An analytical coaxial transmission line model was derived and cross-validated using numerical simulations, resulting in an absolute fractional difference < 5 per cent across the 50–350 MHz frequency range. The formulae were used to construct a per-antenna and per-station gain model for an 8-m RG-58u coaxial transmission line across a range of parameter values with variations in length, impedance mismatch, manufacturing tolerances, and temperature. This study aimed to encapsulate transmission line attenuation and reflection effects that may have an impact on SKA-Low.

The pipeline creates a composite sky model using radio foreground data from GLEAM and Haslam 408 MHz, along with a simulated 21-cm cosmological signal generated using 21cmSPACE. To cover the primary beam of SKA-Low, the 21-cm simulation boxes were large, with a side length of 1.5 cGpc. Instantaneous snapshots of the MWA EOR1 field were simulated using the SKA-Low core across 160 channels from 145 to 165 MHz for each transmission line case. A per-baseline delay power spectrum showing the foregrounds and 21-cm signal was produced for each observation. To quantitatively assess the contaminated k-mode bins, the delay power spectrum was subdivided into the foreground wedge and two k-mode regions of the EoR window: those with suppressed foreground emission, and those dominated by supra-horizon emission.

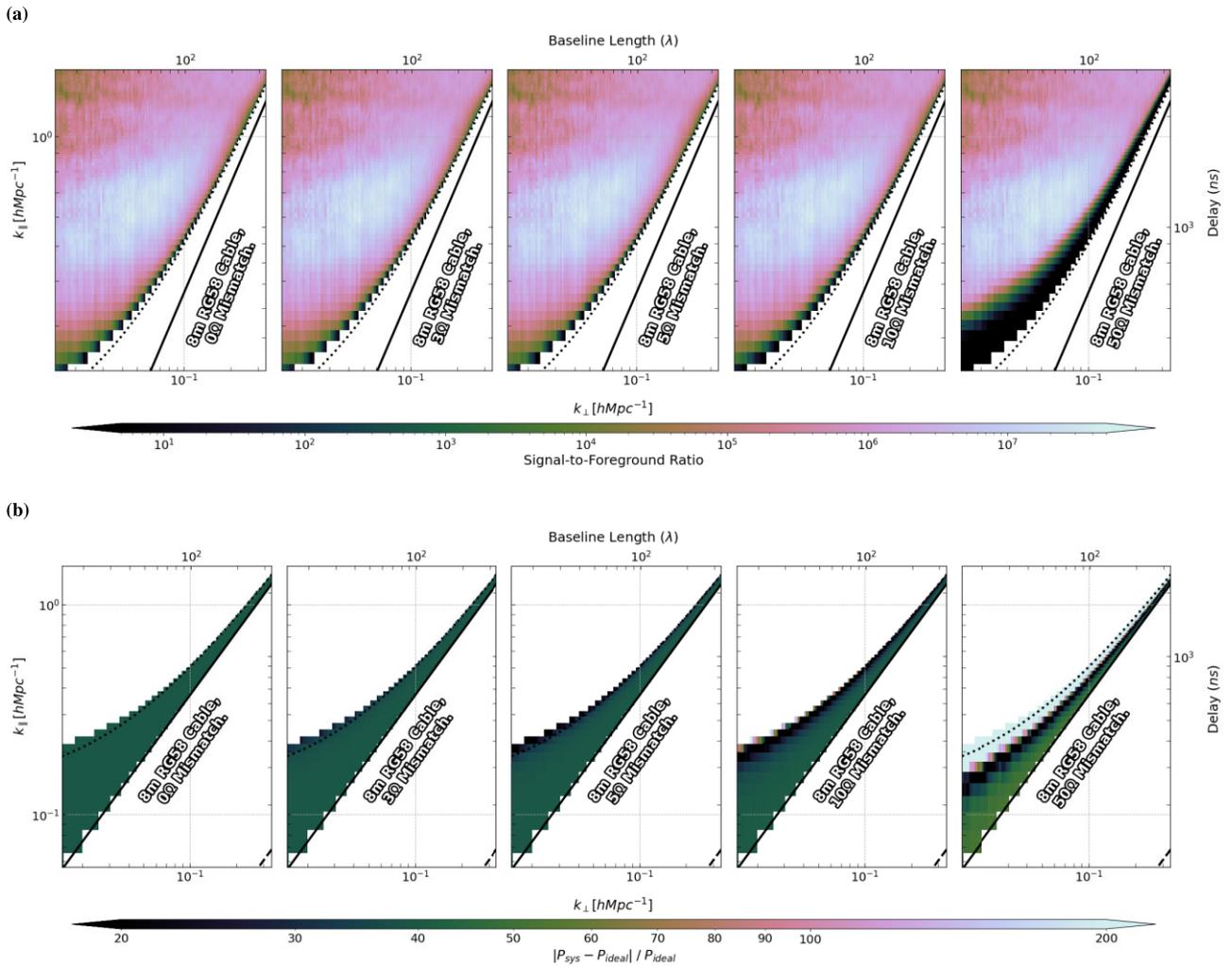

**Figure 11.** A parametric sweep of transmission line impedance mismatch, as outlined in Table 2, illustrating the spectral contamination of (a) the EoR window and (b) the supra-horizon dominated k-modes. The magnitude of the impedance mismatch determines the percentage of power reflected at the transmission line interface and thus the magnitude of the signal-to-foreground minima observed in delay space. For an 8-m line, the peaks are confined below the horizon limit (solid black line) for realistic impedance mismatches of $\leq 10\Omega$. However, cracks in the transmission line resulting in mismatches upwards of $50\Omega$ would broaden the peaks by amplifying their tails with respect to the surrounding foreground emission. In the case of a perfectly matched front-end, a $\sim 37$ per cent absolute fractional difference is observed due to attenuative effects, thus indicating that a successful calibration must account for transmission line reflection and dispersion effects.

Instrumental design typically aims to minimize signal reflections by ensuring components are impedance-matched. Across wide bands, a perfect match is not possible, so transmission line lengths should be selected to avoid placing the reflection peaks at delays corresponding to the EoR window. For short 8- and 15-m cables, an impedance mismatch extending to $10\Omega$ imparts negligible contamination by confining the reflection to k-modes below the EoR window. However, the region of the power spectrum dominated by foregrounds is affected by the cable-length-dependent attenuation. A difference of $\sim 37$ per cent due to this attenuation occurs even with a perfectly impedance-matched 8-m transmission line. Techniques that rely on foreground removal methods should incorporate all cable characteristics to recover the 21-cm signal.

Furthermore, we demonstrate the impact of aliasing for long transmission lines whose delay extends beyond the maximum $k_\parallel$ mode. Thus, we highlight the importance of line attenuation and the inclusion of ferrite attenuation beads, which aim to reduce the reflected power below that of the 21-cm signal. This is observed for a 1-km transmission line where the aliased reflection imparts a negligible effect on the EoR window.

Random fluctuations in transmission line temperature across the array result in an attenuation effect. While this indicates no apparent change in the 21-cm signal-to-foreground ratio, we observe a calculation artefact along the horizon limit buffer. Furthermore, due to this effect being purely attenuative, and the smooth thermal gradient present across the array, amplitude calibration of the beam should correct for this attenuation. In addition, manufacturing tolerances further complicate the calibration process, introducing a noise-like component in the delay power spectrum due to varying station beam side-lobes. These length tolerances also result in the broadening of the reflection peaks.

We conclude that the 7-m coaxial cables used in AAVS2, and proposed for SKA-Low, should not prevent a foreground-avoidance approach from being used to detect the 21-cm signal due to confinement of reflected power in the foreground wedge. Even when using cables of zero length tolerance, the visibility power is attenuated, which modifies the predicted foregrounds; and thus any use of a foreground-removal algorithm must account for this chromatic attenuation to perform a successful calibration. This effect is further complicated by manufacturing standards quoting coaxial cable length tolerances to 1.0 per cent, resulting in unique element gains and therefore station beams, emphasizing the need for a per-antenna, per-station gain correction within the calibration pipeline.

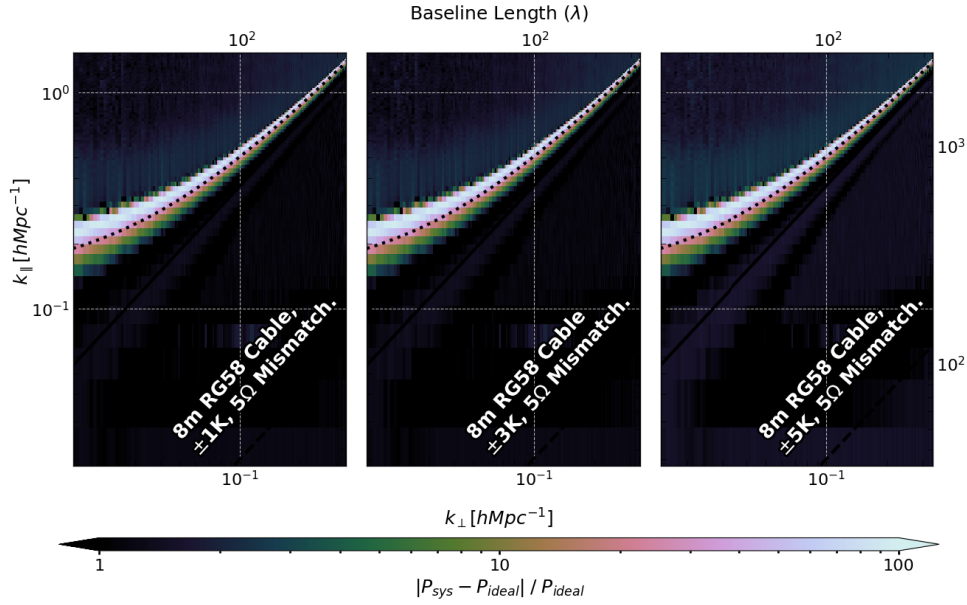

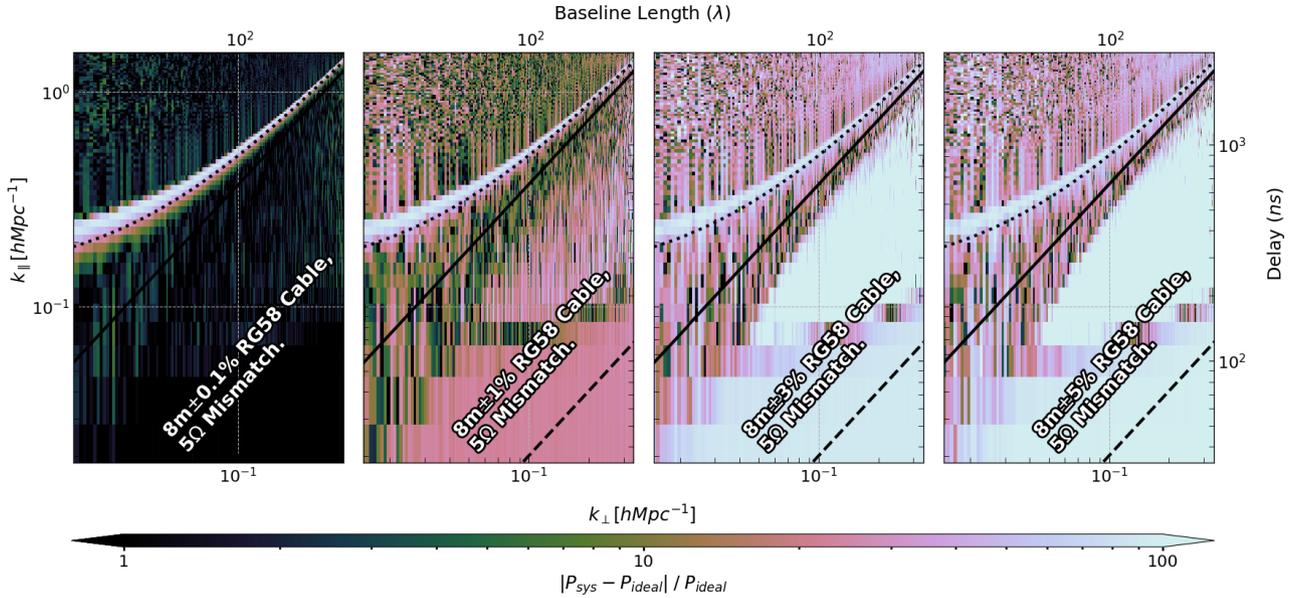

**Figure 12.** A comparison of absolute fractional differences between the reference observation outlined in Table 2 and the inclusion of cable length tolerance and thermal variations also outlined in Table 2. The differences were computed in visibility space and plotted using the per-baseline delay power spectrum method in Section 2.5. Across the bottom panels, we observe structure in the noise streaking vertically, which is superimposed upon the structure found in Fig. 12(a).

## ACKNOWLEDGEMENTS

The authors would like to extend their thanks to the anonymous referee for their insightful comments, which helped to improve the clarity and scientific rigour of this manuscript.

In addition, we would like to thank Clark Huang for initializing an extension of 21cmSPACE to support larger simulation cubes to encompass the SKA-Low primary beam, and Yuki Pan for the early application of OSKAR on the 21cmSPACE simulation cubes to access the signal-to-foreground power ratio in the EoR window.

OSDOH acknowledge the support of ESA and NPL for grant #G109464, 'The Design of Highly Sensitive EM Sensors for Space Applications'. FD, TGJ, DA, and EdLA acknowledge the support of the Science and Technology Facilities Council (STFC) with respective grant numbers ST/X00239X/1, ST/V506606/1, and ST/W00206X/1, and a Rutherford Fellowship. JD acknowledges support from the Boustany Foundation and Cambridge Commonwealth Trust in the form of an Isaac Newton Studentship. AF is supported by a Royal Society University Research Fellowship #180523.

This work was performed using resources provided by the Cambridge Service for Data Driven Discovery (CSD3) operated by the University of Cambridge Research Computing Service (www.csd3.cam.ac.uk), provided by Dell EMC and Intel using Tier-2 funding from the Engineering and Physical Sciences Research Council (capital grant EP/T022159/1), and DiRAC funding from the Science and Technology Facilities Council (www.dirac.ac.uk).

## DATA AVAILABILITY

The full end-to-end pipeline outlined in this work is located on GitHub at https://github.com/oharao/SKA_Power_Spectrum_and_EoR_Window/, alongside installation instructions for the required dependences in the form of a Docker container.

This paper has been typeset from a TeX/LaTeX file prepared by the author.